\documentclass[aps,prb,twocolumn,superscriptaddress,showpacs]{revtex4}
\usepackage{graphicx}
\usepackage{amsmath,amsfonts,amssymb}

\newcommand{\mtimes}{{$\times$}}

\newcommand{\centigrade}{{$^\circ$\textrm{C}}}
\newcommand{\degree}{{$^\circ$}}
\newcommand{\didv}{{\textit{dI/dV}}}

\begin{document}

\title{One-dimensional surface states on a striped Ag thin film with stacking fault arrays}

\author{Takashi Uchihashi}
\email{UCHIHASHI.Takashi@nims.go.jp}
\affiliation{International Center for Materials Nanoarchitectonics, National Institute for Materials Science, 1-1 Namiki, Tsukuba 305-0044, Japan}

\author{Puneet Mishra}
\affiliation{International Center for Materials Nanoarchitectonics, National Institute for Materials Science, 1-1 Namiki, Tsukuba 305-0044, Japan}

\author{Katsuyoshi Kobayashi}
\affiliation{Department of Physics, Faculty of Science, Ochanomizu University, 2-1-1 Otsuka, Bunkyo-ku, Tokyo 112-8610, Japan}

\author{Tomonobu Nakayama}
\affiliation{International Center for Materials Nanoarchitectonics, National Institute for Materials Science, 1-1 Namiki, Tsukuba 305-0044, Japan}
\date{\today}

\begin{abstract}
One-dimensional (1D) stripe structures with a periodicity of 1.3 nm are formed by introduction of stacking fault arrays into a Ag thin film. 
The surface states of such striped Ag thin films are studied using a low temperature scanning tunneling microscope.
Standing waves running in the longitudinal direction and characteristic spectral peaks are observed by differential conductance (\didv) measurements, revealing the presence of 1D states on the surface stripes.
Their formation can be attributed to quantum confinement of Ag(111) surface states into a stripe by stacking faults.
To quantify the degree of confinement, the effective potential barrier at the stacking fault for Ag(111) surface states is estimated from independent measurements.
A single quantum well model with the effective potential barrier can reproduce the main features of \didv\ spectra on stripes, while a Kronig-Penney model fails to do so.
Thus the present system should be viewed as decoupled 1D states on individual stripes rather than as anisotropic 2D Bloch states extending over a stripe array.
\end{abstract}


\maketitle

\section{Introduction}
\label{sect:Introduction}
Growth of a thin metal film on a semiconductor substrate is usually governed by the Volmer-Weber mode or the Stranski-Krastanov mode, resulting in formation of an inhomogeneous and granular film. 
The rapid advancements of nanotechnology, however, has motivated researchers to fabricate epitaxial films on a semiconductor such as silicon with atomic-scale precision.
This was accomplished by what is called two-step growth method \cite{Hoegen_AgSi,Gavioli_AgSi,Jiang_AgSi,Guo_PbQW,Neuhold_AgSiSS,Matsuda_AgSiQWS,Smith_AgGaAs,Jiang_AgGaAs,Yeh_PbSi,Su_PbSi,Ricci_PbSi,Sawa_AgSSDislocation,Tang_AgGe,Uchihashi_AgStripe,Nagamura_AgStripe,Uchihashi_AgSi(111)In,Okuda_AgStripes,Uchihashi_AgSFReflection} utilizing a kinetic path to avoid a route to the thermodynamically stable states.
Thus fabricated ultrathin metal films can possess Shockley surface states \cite{Neuhold_AgSiSS,Jiang_AgGaAs,Sawa_AgSSDislocation,Uchihashi_AgSFReflection} and quantum well states \cite{Neuhold_AgSiSS,Matsuda_AgSiQWS,Jiang_AgGaAs,Nagamura_AgStripe,Okuda_AgStripes} due to the electron confinement in the normal direction, both of which are well-defined two-dimensional (2D) systems.
They offer ideal platforms for studying intricate electronic interactions with the substrate \cite{Yeh_PbSi,Ricci_PbSi,Tang_AgGe} and various quantum effects manifesting themselves,  \textit{e.g.}, in the film growth behaviors \cite{Gavioli_AgSi,Su_PbSi,Ricci_PbSi,Yeh_PbSi,Guo_PbQW} and superconductivity \cite{Guo_PbQW}.
These have been revealed with unprecedented accuracies by surface-sensitive techniques such as scanning tunneling microscopy/spectroscopy (STM/S) \cite{Jiang_AgGaAs,Sawa_AgSSDislocation,Uchihashi_AgSFReflection} and photoemission spectroscopy \cite{Neuhold_AgSiSS,Matsuda_AgSiQWS,Ricci_PbSi,Tang_AgGe,Nagamura_AgStripe,Okuda_AgStripes}.

The growth behaviors of such epitaxial thin films are strongly influenced by a substrate surface \cite{Smith_AgGaAs,Jiang_AgGaAs,Jiang_AgSi}.
Recently, it was found by Uchihashi \textit{et al.} that the atomic structure of a Ag thin film can be periodically modulated using a substrate with one-dimensional surface structures \cite{Uchihashi_AgStripe,Uchihashi_AgSi(111)In}.
The surface of a silicon substrate was decorated with arrays of In atomic chains with a periodicity of 1.33 nm in a self-assembling fashion.
This surface reconstruction, Si(111)-(4\mtimes1)-In \cite{Kraft_InSurfaces,Bunk_In4x1,Nakamura_In4x1,Miwa_In4x1,Abukawa_In4x1,Yeom_In4x1}, plays the role of a geometrical template for the growth of a Ag film and introduces stacking fault (SF) arrays.
Figure 1(a) shows a representative topographic STM image (displaying the height $z$) of such a Ag thin film with a nominal thickness of 22 monolayers (ML).
The overall surface morphology is flat and exposes only a few layers as a result of epitaxial growth.
A careful inspection reveals that the surface is composed of narrow parallel stripes.
The lower-right section of the image displays the derivative of the topographic height ($dz/dx$, $x$: the horizontal coordinate), where parallel stripe arrays are clearly visible.

The atomic structure of these Ag stripes has been explained as follows (see Fig. 1(b)) \cite{Uchihashi_AgStripe}.
The film surface consists of Ag(111) nano-planes with periodic insertion of SF planes at every five layers.
This results in a relatively good matching between the transverse periodicities of the SF array and the In chains on the substrate and stabilizes the film.
Notably, this SF array has been found to significantly affect the bulk electronic states in the film and  to change the quantum well states into those of 1D character or a high anisotropy \cite{Nagamura_AgStripe,Kobayashi_SFBulk,Okuda_AgStripes}.
The SF array should also modify the Shockley surface states of Ag(111) because of their strong electron reflection and quantum confinement effects \cite{Sawa_AgSSDislocation,Uchihashi_AgSFReflection}.
However, the details of such potentially 1D surface states on a Ag film has not been clarified yet.

\begin{figure}
\begin{center}
\includegraphics[width=80mm]{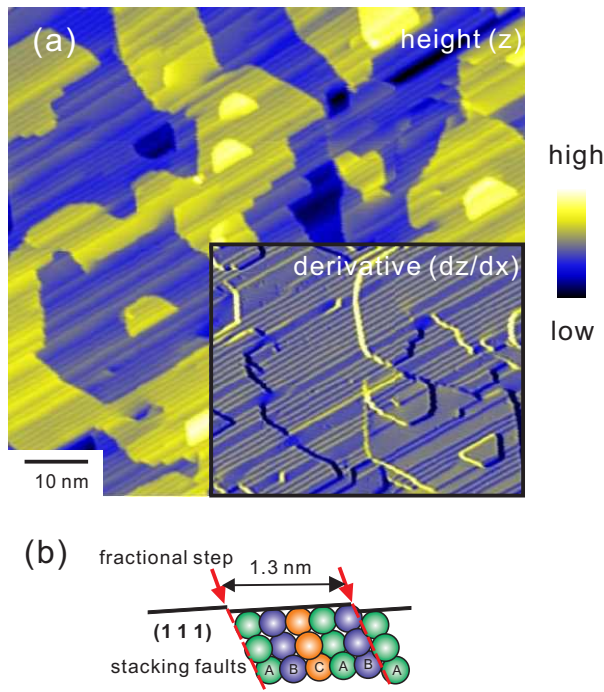}
\end{center}
\caption{(Color online)
(a) Topographic STM image of a striped Ag film with a nominal thickness of 22 ML grown on the Si(111)-(4\mtimes1)-In surface ($V=2$ V, $I=120$ pA). The lower-right section displays the derivative of the topographic height ($dz/dx$, $z$: topographic height, $x$: the horizontal coordinate).
(b) The atomic structural model of the Ag stripes with stacking fault planes.
}
\end{figure}

In this paper, we investigate the surface states of the striped Ag film with SF arrays using a low-temperature (LT) STM.
Differential conductance (\didv) imaging shows clear standing wave patterns along the stripes, the periodicity of which exhibits a systematic energy dependence. 
This allows us to obtain a free-electron-like energy dispersion and to compare it to that of Ag(111) surface states.
\didv\ spectra taken on stripes exhibit a peak around $V\approx 0.3-0.4$ V and their shapes are consistent with the density of states (DOS) in a 1D electron system.
These observations reveal the presence of 1D electronic states on the stripe surface, which is formed by quantum confinement of the Ag(111) surface states within a stripe.
To quantify the degree of confinement, the effective potential barrier at the SF for surface states was estimated from independent spectral measurements on a wider area.
A single quantum well (SQW) model with the above effective potential barrier can reproduce the main features of \didv\ spectra on stripes, while a Kronig-Penney (KP) model fails to do so.
This indicates that the surface states on the striped Ag film should be viewed as decoupled 1D states rather than anisotropic 2D Bloch waves extending over periodic arrays of stripes.

The present paper is organized as follows.
In Section~\ref{sect:Exp}, the experimental methods are explained.
Section~\ref{sect:Results} describes the results on \didv\ measurements and the determination of the effective potential barrier at a SF.
In Section~\ref{sect:Discussion}, we calculate the DOS as a function of energy on a stripe using the two models to compare them to the experimental data.
The summary is given in Section~\ref{sect:Conclusion}.

\section{Experimental Methods}
\label{sect:Exp}
The experiments were performed in an ultrahigh vacuum system equipped with a low temperature STM and a low energy electron diffraction (LEED) optics.
First, a Si(111)-(7\mtimes7) clean surface was prepared by high-temperature flashing up to 1280 \centigrade.
Deposition of a small amount of indium ($\sim$ 1.8 ML) on the surface followed by sample annealing around 340 \centigrade\ for 5 min resulted in formation of a Si(111)-(4\mtimes1)-In surface reconstruction \cite{Kraft_InSurfaces,Yeom_In4x1,Abukawa_In4x1,Bunk_In4x1}.
This was confirmed by LEED and, if necessary, was further observed by STM to optimize its domain sizes and defect density.
Ag was deposited up to a coverage of 22 ML onto the substrate that was cooled down below 100 K, followed by natural annealing to room temperature (RT).
This two-step growth process is crucial for creating a flat epitaxial Ag film with a good crystallinity on a silicon substrate \cite{Hoegen_AgSi,Gavioli_AgSi,Jiang_AgSi,Neuhold_AgSiSS,Matsuda_AgSiQWS}.
As a result of the templating effect of the Si(111)-(4\mtimes1)-In substrate, periodic SF planes were introduced into the Ag film with a mean transverse periodicity of 1.3 nm \cite{Uchihashi_AgStripe}.
The SF planes are terminated on the surface by ``fractional steps'' with a height of 0.078 nm (=1/3 of monatomic step height), as shown in Fig. 1(b).
The most frequently observed terrace bounded by parallel fractional steps has a width of 1.3 nm equal to the periodicity, which is referred to here as \textit{single-unit} stripe.
Occasionally a terrace with a width of 2.6 nm was also observed, which is referred to as \textit{double-unit} stripe.  
Although the surface of a striped Ag film also includes monatomic steps with a height 0.31 nm, they can be easily distinguished from fractional steps by measuring the step height.

All STM spectroscopic measurements and accompanying topographic imaging were performed below 8 K.
W tips prepared by flashing and $\mathrm{Ar^+}$ ion bombardment were further shaped by slight touching to a sample surface. 
A sample bias voltage $V$ was measured relative to the tip, which is converted to the energy level $E$ of a sample state through a relation of $E=eV$.
Here $E$ is measured relative to the Fermi level. 
\didv spectra were acquired by standard lock-in ac detection.
The modulation amplitude $V_{\text{mod}}$ was set to $5-20 \ \text{mV}_{\text{p-p}}$ according to the energy resolution required for a specific measurement.
Since the feedback was stabilized at a relatively high voltage of 1 V, we neglect tip height variations at different locations and thus regard the measured spectrum as being proportional to the sample DOS.
The effect of the energy-dependent tip DOS should be marginal because we concentrate on the empty states of the sample here \cite{Wisendanger_STMTextbook}.
\didv\ images were taken with $V_{\text{mod}}=35-40 \ \text{mV}_{\text{p-p}}$ while scanning in the constant current mode.

\section{Results}
\label{sect:Results}
\subsection{\didv\ imaging on single-unit stripes}
\label{subsect:didvi_mage}
Figure 2(a) shows a typical topographic STM image of parallel stripes running through a monatomic-height island.
The positions of the monatomic and fractional steps are indicated by solid and dashed lines, respectively 
The island includes five stripes labeled by (A)-(E), among which (B) and (C) are single-unit stripes bounded by fractional steps.
Because of the presence of the island edge, standing waves are expected to arise if electrons can move freely along the stripes \cite{Crommie_QuantumCorral,Hasegawa_AuSW,Avouris_AuQSE,Li_AgSW,Li_AgIsland,Burgi_AgResonator,Pivetta_RipplesAgSS}.
This is indeed observed in \didv\ images where the signal is periodically modulated along the stripes, an example of which is shown in Fig. 2(b).
Each stripe has different amplitudes of modulation, and almost no modulation in \didv\ signal is visible on the fractional steps.
This suggests that the electronic states on stripes are of 1D character, decoupled from each other by the presence of fractional steps.
Figure 3(c) plots a series of \didv\ signal taken along the middle of the stripe (B) with different sample voltages $V$ from 0.38 to 0.68 V.
The periodicity of the modulation decreases with increasing $V$, which allows us to attribute the phenomenon to surface electronic standing waves \cite{Crommie_QuantumCorral,Hasegawa_AuSW,Avouris_AuQSE,Li_AgSW,Li_AgIsland,Burgi_AgResonator,Pivetta_RipplesAgSS}.

\begin{figure}
\begin{center}
\includegraphics[width=84mm]{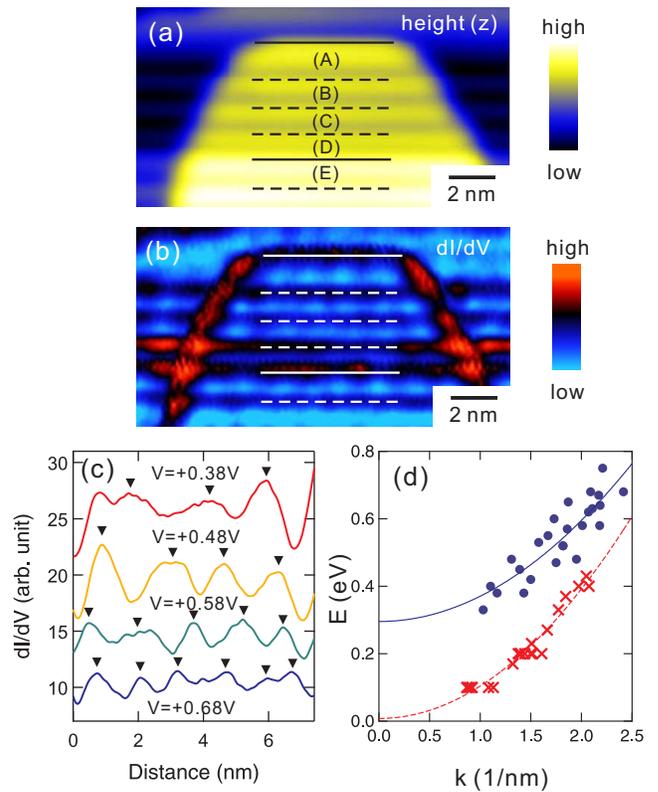}
\end{center}
\caption{(Color online)
(a) Topographic STM image of parallel stripes running through a monatomic-height island ($V=0.6$ V, $I=120$ pA). The positions of the monatomic and fractional steps are indicated by solid and dashed lines, respectively. The narrow striped regions are labeled by (A)-(E), among which (B) and (C) are single-unit stripes.
(b) \didv\ image on the same area as in (a) showing the periodically modulated signal along the stripes ($V=0.58$ V, $I=120$ pA).
(c) Series of \didv\ signal taken along the middle of a stripe (B) with different sample voltages $V$ from 0.38 to 0.68 V. The data are offset along the ordinate for clarity.
(d) Energy dispersion of the surface states obtained from the standing wave observations. Closed circles and crosses are data taken on single-unit stripes and a flat area of a Ag(111) film epitaxially grown on Si(111), respectively.
The solid and dashed lines show the fitting results using Eq.~(\ref{eq:free_electron_dispersion}).
}
\end{figure}

The voltage-dependent \didv\ imaging on a single-unit stripe as performed above was repeated to obtain the energy dispersion of the surface states.
The wavenumber $k$ of the \textit{wavefunciton} probed at an energy $E=eV$ is determined through a relation $k=\pi/\lambda$, where $\lambda$ is the wavelength of the \textit{standing wave}.
The results are summarized in Fig. 2(d) (solid circles).
Because $\lambda$ is not strictly reproduced at the same $E$ for different stripes, the data are distributed along the ordinate within the range of $\approx 0.15$ eV; however, they show a clear trend of increasing $E$ with increasing $k$.
The data were fitted using a free-electron-like dispersion relation:
\begin{equation}
 E=E_0+\frac{\hbar^2 k^2}{2m^*}  \label{eq:free_electron_dispersion}
\end{equation} 
 where $E_0$ is the onset energy of the surface band and $m^*$ its effective mass.
The fitting analysis gives $E_0 = 0.30 \pm 0.03 \ \mathrm{eV} (\equiv E_\mathrm{0,stripe})$ and $m^* =0.51 \pm 0.06 \ m_0 (\equiv m^*_\mathrm{stripe})$, where $m_0$ is the free electron mass.
For comparison, a similar experiment was performed on a flat surface (without stripes) of an epitaxially grown Ag(111) film on S(111) (crosses in Fig. 2(d)). 
The same fitting analysis using Eq.~(\ref{eq:free_electron_dispersion}) gives $E_0= 0.01 \pm 0.01 \mathrm{eV}(\equiv E_\mathrm{0,flat})$  and $m^* =(0.40 \pm 0.02) m_0 (\equiv m^*_\mathrm{flat})$, the result of which is plotted in Fig. 2(d) as the dashed line.
Considering that the two energy dispersions are similar, we can conjecture that these 1D surface states originate from Ag(111) surface states.
Although 1D standing waves have been reported on a Ag(111) surface decorated with biomolecular gratings \cite{Pennec_SupramolecularGrating}, their existence on such a narrow stripe has not been clear so far.
We note that, in the case of 1D standing waves on Si(001) surfaces \cite{Yokoyama_Si(001)SS,Sagisaka_Si(001)}, the electrons are confined within a dimer row due to the local dangling bonds, which is a different mechanism from the one for the present system.

The large difference  $\sim 0.3$ eV between $E_{0,\mathrm{stripe}}$ and $E_{0,\mathrm{flat}}$ is attributed to a quantum confinement in the transverse direction as corroborated later \cite{Morgenstern_AgSSdep,Ortega_1Dvs2DAgSS,Mugarza_AuResonator,Mugarza_VicinalSurfaces,Burgi_AgResonator,Avouris_AuQSE}.
It is also worth noting that the onset energy of the Ag flat surface $E_{0,\mathrm{flat}}$ is increased from the value for bulk samples  ($E_\mathrm{0}=-0.065 \sim -0.063$ eV) although the effective mass is nearly the same ($m^*=0.40 \sim 0.42 m_0$ for bulk) \cite{Li_AgSW,Burgi_AgResonator,Pivetta_RipplesAgSS}.
This has been attributed to a tensile strain in the film due to a lattice mismatch between the epitaxially grown Ag film and a silicon substrate \cite{Neuhold_AgSiSS}.
A small tensile strain of 0.95 \% is sufficient to shift up the Ag(111) surface states rigidly in energy by 0.15 eV.
The effects of the quantum confinement and the tensile strain on the onset energy $E_0$ will be discussed in Section~\ref{sect:Discussion}.

\subsection{\didv\ spectra on single-unit stripes}
\label{subsect:didvi_spectra_single}
To study the electronic states on single-unit stripes in more detail, \didv\ spectra were measured on different locations on Ag stripes.
Figure 3(a) shows an STM image of the central part of a four-stripe array with a length of $\sim 15$ nm.
The topographic height $z$ averaged along the stripe direction is displayed below the image.
The dashed lines indicate the locations of fractional steps, which were determined from the steepest slope in $z$ \cite{Li_AgSW}.
First the STM tip was moved along the dotted line indicated by \#1 and \didv\ spectra were taken at the seven sites (shown by the crosses). 
It was then moved onto the locations indicated by \#2 -- \#5 successively and the same measurement was repeated (the spectral sites are not explicitly shown).
The data were averaged for each line and are summarized in Fig. 3(b) (raw data: crosses, average: solid lines).

\begin{figure}
\begin{center}
\includegraphics[width=84mm]{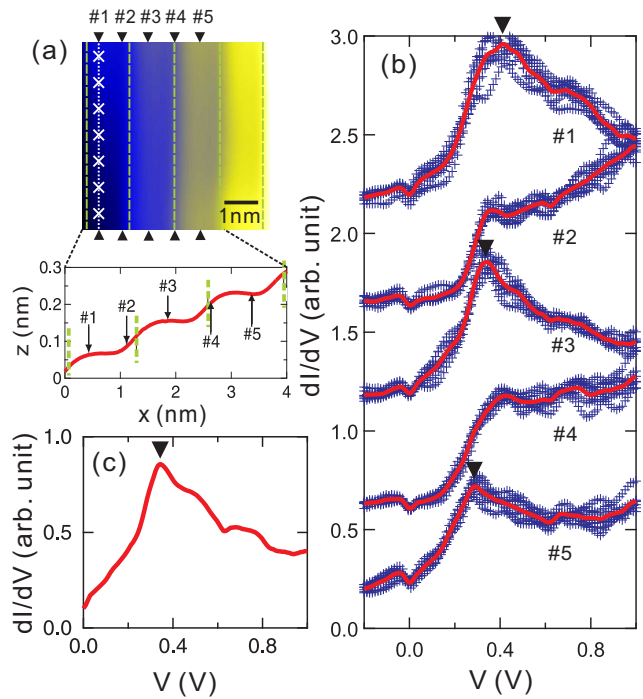}
\end{center}
\caption{(Color online)
(a) Topographic STM image of the central part of an array of four singe-unit stripes ($V=0.5$ V, $I=120$ pA). The dashed lines indicate the locations of fractional steps. The crosses on the dotted line indicated by \#1 show the spectral sites for \didv\ measurements. The same measurements were performed at the locations indicated by \#2 -- \#5 (the spectral sites are not explicitly shown).
The lower panel shows the topographic height $z$ averaged along the longitudinal direction of the stripes.
(b) Raw spectra taken on the locations indicated by the \#1 -- \#5 (crosses) and the averaged spectra for each data set (solid lines). The data are offset along the ordinate for clarity.
(c) Averaged spectra taken in the middle of an isolated single-unit stripe.
}
\end{figure}

The raw spectra are reproduced reasonably well for each data set, which is consistent with the presence of 1D electronic states.
Small variations in spectral shapes are probably due to atomic-scale defects on fractional steps that are not resolved by the topographic imaging.
As expected, the data taken near the middle of the stripes (\#1, \#3, \#5) exhibit clear peaks, but they appear at slightly different voltages of 0.41, 0.33, 0.28 V (indicated by triangles).
This variation is consistent with the distribution of the energy dispersion data obtained from standing wave images in the preceding section.
Note that the energy increase $\delta E$ due to quantum confinement in the longitudinal direction is negligible since $\delta E$ is estimated to be 5 meV at most for the stripe length of 15 nm. 
We also note that the slope of the each peak at the lower energy side is steeper than that at the higher energy side.
This feature is characteristic of the DOS of a sub-band formed in a 1D system; the peak is due to the energy quantization in the transverse direction and the gradual decrease above the peak is due to the free electron motion along the stripes.
Since the surface states exist only for $E>0$ (empty states) due the strain effect mentioned earlier, the observed peaks correspond to the lowest sub-band.
As expected, the peak structure is suppressed for spectra taken near the fractional steps (\#2, \#4).
However, the fact that the peak does not completely vanish at the boundaries indicates that electron confinement within a stripe is not perfect.

For comparison, \didv\ spectra were taken on an \textit{isolated} single-unit stripe (not in an array).
Figure 3(c) shows the average of 10 raw data taken in the middle of the stripe.
The peak positions ($V=0.34$ V) and the shapes of the spectra are very similar to those observed on the stripe array (\#1, \#3, \#5 in Fig. 3(b)).
This means that the electronic states in a stripe array are basically the same as those of an isolated stripe, which again suggests that stripes in an array are decoupled.
We will return to this issue in Section~\ref{sect:Discussion}.

\subsection{\didv\ spectra on double-unit stripes}
\label{subsect:didvi_spectra_double}
Since single-unit stripes as studied above are very narrow, their electronic states may be sensitive to residual defects on fractional steps as mentioned above.
In addition, precise identification of the spectral cites is not easy due to the finite size of the probing tip and the presence of steps.
For quantitative analysis, therefore, we measured \didv\ spectra on a double-unit stripe where the above difficulties were mitigated to a considerable degree.
This allows us to extract information on the effective potential barrier at the fractional steps and to perform model calculations for electronic states on single-unit stripes.

\begin{figure}
\begin{center}
\includegraphics[width=84mm]{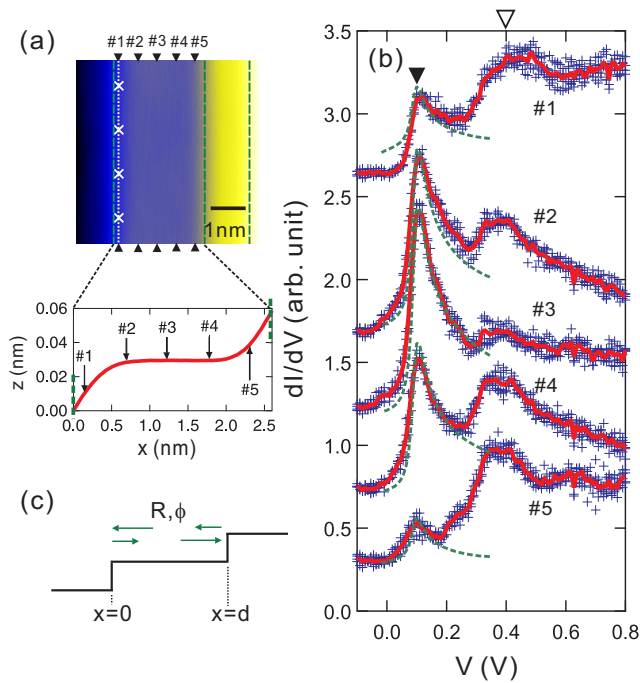}
\end{center}
\caption{(Color online)
(a) Topographic STM image of an area including a double-unit stripe at center and single-unit stripes at sides ($V=-2$ V, $I=120$ pA). The crosses on the dotted line indicated by \#1 show the spectral site where \didv\ measurements were taken. The same measurements were performed at the locations indicated by \#2 -- \#5 (the spectral sites are not explicitly shown).
The lower panel shows the topographic height $z$ averaged in the longitudinal direction of the stripes.
(b) Raw spectra taken on lines the \#1 -- \#5 (crosses) and the averaged spectra for individual locations (solid lines). The dashed lines show the fitting results using Eq.~(\ref{eq:Fabry-Perrot}). The data are offset along the ordinate for clarity.
(c) Schematic diagram of the Fabry-Perrot resonator model for determining the reflection amplitude $R$ and the phase shift $\phi$ at the fractional step.
}
\end{figure}

Figure 4(a) shows an topographic STM image of an area including a double-unit stripe with a length of 18 nm and single-unit stripes at sides.
The topographic height $z$ averaged in the longitudinal direction is plotted below the image.
The locations of fractional steps were determined from the steepest slope in $z$ as previously and are indicated by dashed lines in the figures.
Following the procedure described in the preceding section, five longitudinal lines \#1--\#5 were set (the line and the spectral site are not shown for \#2--\#5) and four \didv\ spectra were taken for each line and averaged.
The result is summarized in Fig. 4(b) (raw data: crosses, average: solid lines). First we note that the raw spectra are reproduced very well for each line, suggesting the negligible effects of residual defects.
The energy increase $\delta E$ due to a longitudinal quantum confinement is at most 3 meV for the stripe length of 18 nm and is negligible.
The sharp peak at $V= 0.11$ V (shown by the solid triangle) is attributed to the lowest sub-band as in the case of single-unit stripes.
As expected, it is enhanced in the middle of the double-unit stripe (\#3) and is suppressed near the boundaries (\#1,\#5).
The rather broad peak around $V=0.4$ V (shown by the open triangle) may be attributed to the second lowest sub-band, but since it is enhanced around boundaries (\#1,\#5), the influence of neighboring single-unit stripes cannot be ignored.
Note that the spectra taken on single-unit stripes have peaks around this energy region as shown above. 
We thus restrict ourselves to the lowest energy peak at $V= 0.11$ V for further analysis.

The local density of states (LDOS) of a stripe bounded by parallel steps can be calculated using a Fabry-Perrot resonator model from a 2D surface state band, as demonstrated by B\"urgi \textit{et al} \cite{Burgi_AgResonator}.
Since the \didv\ signal is proportional to the corresponding LDOS, it is given for an energy $E$ and a transverse location $x$ by the following equations:
\begin{widetext}
\begin{eqnarray}
\didv(E,x) &=& A \Big[ C_\mathrm{off}+ \int^k_0 dq \frac{1}{\sqrt{k^2-q^2}} \frac{1}{1+R^4-2R^2\cos(2qd+2\phi)} \nonumber \\
& & \times 2(1-R^2)\left\{ 1+ R^2+ R \cos(2q(x-d)-\phi) + R \cos(2qx+\phi) \right\} \Big],  \\  \label{eq:Fabry-Perrot}
k&=&\sqrt{2m^* (E-E_0)/\hbar^2}.
\end{eqnarray}
\end{widetext}
Here $A$ is a parameter representing the signal intensity, $R$ the reflection amplitude at the steps, $\phi$ the phase shift at the reflection, $d$ the width of the stripe (for the schematic configuration, see Fig. 4(c)).
$E_0$ and $m^*$ are the onset energy and the effective mass of the 2D surface state band, respectively.
$C_\mathrm{off}$ was included as a constant offset to account for the bulk contribution to the surface DOS.
For simplicity, $R$ and $\phi$ are assumed to be identical for the ascending and descending steps here.
The treatment on $R$ can be rationalized since $R$ does not show a clear difference for the two types of steps \cite{Uchihashi_AgSFReflection}.
To compensate the assumption on $\phi$, the locations of spectra were allowed to be shifted uniformly by $\delta x$.
This additional parameter can absorb the error due to the above assumption into an apparent shift of the step locations, which can be caused by the finite tip radius.
$C_\mathrm{off}=0.3$ was determined from independent spectral measurements on a flat area of a surface and $d=2.6$ nm was determined from the topographic measurement.
$m^*=0.4 m_0$ was adopted from literatures on the Ag(111) surface states \cite{Li_AgSW,Burgi_AgResonator}.
These considerations leave $A$, $R$, $\phi$, $E_0$, $\delta x$ as fitting parameters. 
A reasonably good agreement with the experiment was obtained for all spectra \#1--\#5 (see dashed lines in Fig. 4(b)), giving $R=0.85$ and $\phi=-0.87 \pi$.
Since the analysis was performed for the lowest peak in Fig. 4(b), these values are valid  for an electron energy $E \sim 0.1$ eV.

\section{Model calculation and discussion}
\label{sect:Discussion}
\subsection{Determination of the effective potential barrier}
\label{subsect:potential_barrier}
In this section, we calculate the DOS in isolated and arrayed single-unit stripes to be compared to the experimentally obtained spectra. 
In the preceding section, we have obtained the reflection amplitude $R$ and the phase shift $\phi$ at the fractional step, but the energy range where they are applicable is limited to $E \sim 0.1$ eV.
To calculate the spectra for a wider energy range, we extract the effective potential barrier and perform a model calculation according to the prescription by H{\"o}rmandinger and Pendry \cite{Pendry_1DScatterer}.

For simplicity, the potential barrier at the fractional step is assumed to be that of rectangular shape with constant potential $W$ within the barrier and a width of $a$ (see the inset of Fig. 5(b)):
\begin{eqnarray}
V(x)=
\left\{
	\begin{array}{rl}
	W, & \mathrm{for} -a/2 < x < a/2 \\
	0, & \mathrm{elsewhere.}
	\end{array}
\right.
\end{eqnarray}
Because the potential barrier should be concentrated at a very narrow region along the fractional step, we consider the limit of $a \rightarrow 0$ while the product $Wa$ is kept constant, \textit{i.e.}, a $\delta$-function type potential. 
By solving the Sch\"odinger equation, one can obtain the reflection coefficient $r$ at an energy $E$ as follows \cite{Pendry_1DScatterer}:
\begin{eqnarray}
r &=& \frac{P-P^{-1}}{P\rho-(P\rho)^{-1}}e^{-iKa} \label{eq:r},\\ 
P &=& e^{iQa} \label{eq:P}, \\
\rho &=& \frac{K-Q}{K+Q}, \label{eq:rho} \\
K&=&\sqrt{2m^*E/\hbar^2}, \label{eq:K} \\
Q&=&\sqrt{2m^*(E-W)/\hbar^2}, \label{eq:Q}
\end{eqnarray}
where $K$ and $Q$ are the wave numbers outside and inside the potential barrier, respectively.
$E$ is measured from the band onset $E_0$ and $m^*$ is the effective mass of the band ($m^*=0.4 m_0$ is adopted as earlier). 
The reflection amplitude $R$ and the phase shift $\phi$ at the potential barrier is given by $R=|r|$ and $\phi=\arg(r)$.
In the present analysis, $W$ is taken to be complex to account for the fact that electrons can be scattered into the bulk states at the fractional step. 
The requirement of complex $W$ is also necessary to treat $R$ and $\phi$ independently.
Reversely, $W$ can be determined if $R$ and $\phi$ are given for a certain $E$.
From $R=0.85, \phi=-0.87 \pi$ for $E=0.1$ eV determined in Section~\ref{subsect:didvi_spectra_double}, $Wa=0.41-0.07i \ \mathrm{eV nm}$ is obtained for $a\rightarrow 0$. 
The real and imaginary parts of $Wa$ change only by -7 \% and -16 \%, respectively, when $a$ is increased up to 0.1 nm.

\begin{figure}
\begin{center}
\includegraphics[width=60mm]{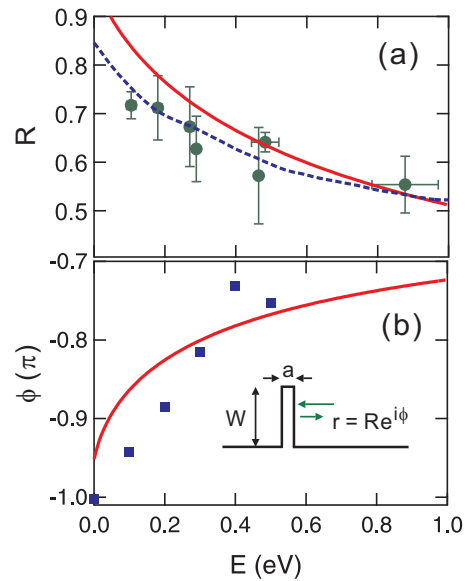}
\end{center}
\caption{(Color online)
(a) Reflection amplitude at the fractional steps $R$ calculated using Eq.~(\ref{eq:r}) based on the effective potential barrier (solid line). The solid circles and the dashed line show the previous experimental result and the tight-binding calculation on $R$, respectively \cite{Uchihashi_AgSFReflection}.
(b) Phase shift at the reflection $\phi$ calculated with the same method as in (a). The closed squares show the result of the tight-binding calculation. Inset: Schematic diagram of the potential barrier used in the model calculation.
}
\end{figure}

Now that $Wa$ has been determined, reflection amplitude $R$ and phase shift $\phi$ can be calculated using Eq.~(\ref{eq:r}) as a function of $E$.
The results are plotted as solid lines in Fig. 5(a)(b), together with the values determined by independent methods  \cite{Uchihashi_AgSFReflection}.
The solid circles represent $R$ (for both descending and ascending fractional steps), which were determined from electron lifetimes in quantum boxes bounded by fractional steps.
The dashed lines in Fig. 5(a) and the closed squares in Fig. 5(b) show values obtained by a theoretical calculation based on a tight-binding method.
Here the atomic structure of the fractional step shown in Fig. 1(b) was adopted and the tight-binding parameters were determined to reproduce a band structure of a Ag(111) thin film obtained by a density-functional method (for details, see Refs~\onlinecite{Kobayashi_SFBulk,Uchihashi_AgSFReflection}).
We note that $R$ and $\phi$ determined here are in reasonable agreement with the independent experiment and calculation, supporting the present analysis.
We therefore adopt the effective potential barrier $Wa=0.41-0.07i \ \mathrm{eV nm}$ in the following analysis.

\subsection{Calculation of the density of states on single-unit stripes}
\label{subsect:calc_stripe}
Having determined the effective potential barrier of the fractional step, we can now calculate the DOS of single-unit stripes.
We consider here two simple models: single quantum well (SQW) and Kronig-Penney (KP) models.
The SQW model consists of two identical potential barriers with a height of $W$ and a width of $a$ and of a well region with a width of $d-a$ (see Fig. 6(a)).
The KP model consists of the potential barriers with the same $W$ and $a$, which are infinitely repeated with a periodicity of $d$ (see Fig. 6(b)).
In both cases, the potential outside the barriers is constant ($=E_0$) and the electron effective mass are set to be $m^*$.
We consider the limit $a\rightarrow 0$ again.

The eigen energies $E_n (n=0,1)$ of the SQW model can be analytically calculated by matching the wavefunctions at the boundaries.
$E_n$ are obtained by solving the following equation:
\begin{eqnarray}
 e^{i2Qa}\rho \frac{\rho\pm e^{iK(d-a)}}{\rho e^{iK(d-a)} \pm 1}=\pm 1
\end{eqnarray}
where $\rho$, $K$, and $Q$ are defined by Eqs.~(\ref{eq:rho})-(\ref{eq:Q}) and the signs $-,+$ corresponds to $n= 0,1$, respectively.
For $Wa=0.41-0.07i$ and $d=1.3$ nm, one obtains $E_0=0.32-0.08i$ eV and $E_1=1.44-0.43i$ eV.
The eigen energies $E_n$ are complex because of finite lifetimes of the electron due to escaping from the well region.
Since the second lowest level ($n=1$) is already quite high and has a large imaginary part (i.e., the energy level is broadened), we do not consider higher levels.
Based on these results, the DOS in the well is calculated using a Lorentzian function:
\begin{equation}
\rho_\mathrm{T}(E)= \frac{1}{\pi} \displaystyle \sum_n \frac{\mathrm{Im}(E_n)}{(E-\mathrm{Re}(E_n))^2+\mathrm{Im}(E_n)^2}. \label{eq:Lorentzian}
\end{equation}
The subscript ``T'' indicates that only the transverse motion of electrons are considered here. 
The actual stripe on the Ag film has an additional degree of freedom in the longitudinal direction.
The DOS including this contribution is calculated as follows:
\begin{equation}
\rho_\mathrm{T+L}(E)= \int_{0}^{E}  \rho_\mathrm{T}(E_x) \frac{1}{\sqrt{E-E_x}} dE_x, \label{eq:convolution}
\end{equation}
where the subscript ``T+L'' indicates that both the transverse and longitudinal motions are considered.
The results for $\rho_\mathrm{T}(E)$ and $\rho_\mathrm{T+L}(E)$ are shown as solid lines in Fig. 6(c)(d). 
Since $E_0$ is close to 0 eV for the present Ag(111) surface states, $\rho_\mathrm{T+L}(E)$ can be directly compared to the experimental results.
We note that the DOS calculated this way is not locally resolved; however, since the $n=0$ level (the lowest sub-band) is dominant here, they can be compared to the spectra taken in the middle of the stripe (\#1, \#3, \#5 in Fig. 3(b) and Fig. 3(c)).
The line shape and the peak position ($E=0.37$ eV) reproduces the experiment at least qualitatively.
In addition, the lowest quantized level in the transverse direction $\mathrm{Re}(E_0)=0.32$ eV (equal to the peak energy in $\rho_\mathrm{T}(E)$) is consistent with the band onset $E_\mathrm{0,stripe} = 0.30 \pm 0.03 \ \mathrm{eV}$ found in Section~\ref{subsect:didvi_mage}.

\begin{figure}
\begin{center}
\includegraphics[width=70mm]{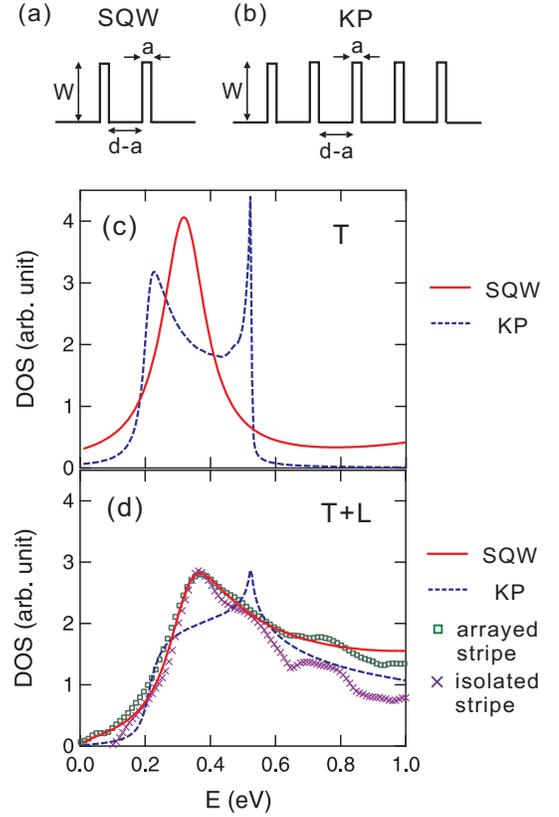}
\end{center}
\caption{(Color online)
(a),(b) Schematic diagrams for (a) the single quantum well (SQW) and  (b) the Kronig-Penney (KP) models. 
(c) DOS for the transverse direction calculated for the SQW and KP models.
(d) DOS including the longitudinal degree of freedom calculated for the SQW and KP models.
Averaged spectra taken on the arrayed and isolated stripes are also shown as open squares and crosses, respectively.
}
\end{figure}

To compare them in more detail, the spectra on the arrayed stripes (\#1, \#3, \#5 of Fig. 3(b)) are averaged and displayed as open squares in Fig. 6(d).
Before averaging, each spectrum was normalized by the peak intensity and was rigidly shifted in energy in such a way that the peak positions were aligned.
This is a reasonable treatment if we assume that the deviation of the peak position is only due to misalignment of the transversely quantized levels as discussed later.
To account for the bulk contribution, a constant value was subtracted from the averaged spectra and then the peak intensity was adjusted to that of the calculated DOS.
The spectrum on the isolated stripe (Fig. 3(c)) was processed in a similar way and is displayed as crosses in Fig. 6(d). 
Evidently, the averaged spectra on the stripes are in good agreement with calculated DOS except at the high energy region.
The deviation may be explained by a site-dependent contribution of the second lowest sub-band, which has a node at the center of the stripe.

Considering that the stripes are usually in the form of an array, the KP model might be more appropriate to describe the DOS in stripes.
To investigate this possibility, we also performed a similar calculation using the KP model.
In this case, the Bloch wave vector $k$ is determined by the following equations \cite{Pendry_1DScatterer}:
\begin{eqnarray}
\cos(kd)&=&\frac{1}{2t}\left[(t^2-r^2)e^{iKd}+e^{-iKd}\right], \label{eq:k} \\
t&=&\frac{\rho-\rho^{-1}}{P\rho-(P\rho)^{-1}}e^{-iKa} \label{eq:t}
\end{eqnarray}
where $t$ is the transmission coefficient of the barrier, and $r$, $P$, $\rho$, and $K$ are given by Eqs.~(\ref{eq:r})-(\ref{eq:K}).
A periodic potential such as the one in the KP model induces energy gaps where $k$ becomes complex.
Energy $E$ permitted for band formation is determined by the condition of real $k$, \textit{i.e.}, $\mathrm{Im} [\cos(ka)]=0$.
If $W$ is complex, $E$  also generally becomes complex.
For the transverse direction, $\rho_\mathrm{T}(E)$  is calculated by Eq.~(\ref{eq:Lorentzian}) where the summation is replaced by an integral.
$\rho_\mathrm{T+L}(E)$ including the longitudinal degree of freedom is obtained by the Eq.~(\ref{eq:convolution}) as previously. 
The result is shown as the dashed lines in Fig. 6(c)(d).
Because of the band formation, the peak found in the SQW model is spread and, for $\rho_\mathrm{T+L}(E)$, only a small peak is visible near the upper edge of the band.
This spectral shape is apparently different from the experimental data (\#1, \#3, \#5 in Fig. 3(b) and Fig. 3(c)).
Although the actual stripes are not in an infinite array, they should exhibit qualitatively same features even for a low number of the array.
We hence conclude that the KP model is not appropriate to describe the present stripe array.

\subsection{Discussions}
\label{subsect:discussions}
The conclusion in the preceding section indicates that the surface states on the striped Ag film should be viewed as decoupled 1D states rather than as anisotropic 2D Bloch states.
This may be explained in terms of energy level misalignment among stripes as follows.
As mentioned earlier, the surface states of stripes have energy levels that are misaligned at most by $\sim 0.15$ eV.
Assuming that all stripes have the same electronic states in the longitudinal direction, this is ascribed to a variation in the transversely quantized levels.
For the tight-binding model, the transfer energy $\gamma$ between the neighboring states is given by $\gamma=W_\mathrm{band}/4$, where $W_\mathrm{band}$ is the band width \cite{Ziman_Textbook}.
$W_\mathrm{band}=0.31$ eV is estimated from the KP model calculation in Fig. 6(c), which gives $\gamma=0.077$ eV.
Since this is comparable to or even smaller than the level misalignment mentioned above, it can prevent individual levels from forming a band, making them more independent.
The origin of the level misalignment can be attributed to spatially inhomogeneous strains in the film \cite{Neuhold_AgSiSS,Uchihashi_AgSFReflection}.
The strain in a Ag(111) film  epitaxially grown on Si(111) is about -1\% \cite{Neuhold_AgSiSS}, but the mismatch between the transverse periodicities of Ag stripes and the Si(111)-(4\mtimes1)-In is -3.8 \% \cite{Uchihashi_AgStripe}.
As mentioned above, a tensile strain of 0.95 \% is sufficient to increase the energy levels of the Ag(111) surface states by 0.15 eV \cite{Neuhold_AgSiSS}.
The enhanced tensile strain due to this mismatch can be locally relaxed by annihilation of a SF plane.
Since this relaxation is expected to occur rather randomly, the strain in the film and, consequently, the surface state levels should become spatially inhomogeneous.

A possible important role of electron scattering from surface to bulk states is worth mentioning.
In the present model calculation, the probability of scattering to bulk states is estimated to be 14 \% at $E=0.3$ eV, which is significantly smaller than the transmission probability of 35 \% at the same energy.
However, if the scattering to bulk states is sufficiently stronger than the transmission to the neighboring stripes, the Bloch-type 2D band is not formed even if the energy levels are aligned; consequently, the surface states on the stripes remain independent.
This is actually the case for the surface states bounded by \textit{monatomic steps} \cite{Mugarza_VicinalSurfaces,Burgi_AgResonator}, where scattering to bulk states is dominant.
In that scenario the effective potential barrier adopted here is not valid and the KP model calculation should be replaced by a theory that takes into account the actual surface and bulk electronic states.
In contrast, our approach should be valid for the SQW model because the energy level calculation involves only electron reflection processes, regardless of the importance of the scattering to bulk states.

Finally, we compare the present result with another form of 1D electronic states on noble metal surfaces.
Ortega \textit{et al.} studied the surface states on a vicinal Au(111) with a periodic array of \textit{monatomic steps} by  angle-resolve photoemission spectroscopy (ARPES) \cite{Mugarza_AuResonator,Ortega_1Dvs2DAgSS,Mugarza_VicinalSurfaces}.
They revealed that the surface states were well confined within a stripe-shaped terrace, featuring the 1D character as long as the terrace width $d$ is equal to or larger than  3.8 nm.
However, as $d$ decreased down to $1.2$ nm, the energy levels of the surface states became broader and their dimensionality changed from 1D to 2D.
This was explained as a consequence of an increase in the surface miscut angle $\alpha$.
As $\alpha$ increases beyond a critical value of $\alpha_c=10.2$\degree\ for Au(111), the energy gap of surface-projected bulk states disappears in the momentum space and the surfaces states become mixed with bulk states.
Since they are not localized on the surface anymore, the effective potential barrier of the step is reduced and the surface states (strictly speaking, surface resonances) extends over the steps as 2D Bloch states \cite{Mugarza_VicinalSurfaces,Ortega_1Dvs2DAgSS}.
At the same time, a level broadening occurs due to mixing with bulk states.

In contrast, single-unit stripes studied here have a very narrow width of $d=1.3$ nm, but they still retain 1D character.
The level broadening $\delta E$ of the transverse quantization is also small.
Considering the good agreement between the SQW model calculation and the experimental spectra shown in Fig. 6(d), $\delta E$ at the lowest sub-band is estimated to be $2\mathrm{Im}(E_0)=0.16$ eV.
This is much smaller than $\delta E= 0.330$ eV for the lowest state of the Au(111) vicinal surface with $d=1.2$ nm, and is comparable to $\delta E= 0.143$ eV for the one with $d=3.8$ nm \cite{Mugarza_VicinalSurfaces}.
The reason for this difference can be two-fold.
First, the present striped surface is tilted only by 3.4\degree\ due to a small step height, which is equivalent to 1/3 of the monatomic step height.
This is sufficiently smaller than the critical angle $\alpha_c=7.3$\degree\ for the Ag(111) surface \cite{Mugarza_VicinalSurfaces}.
Second, in contrast to the monatomic step, the fractional step of the stripe has a SF plane continuing to the bulk region.
This should help to confine the surface states within a stripe even if the surface states penetrate deeper and approaches surface resonance states \cite{Uchihashi_AgSFReflection}.

\section{Conclusion}
\label{sect:Conclusion}
We have shown that 1D electronic states originating from the Ag(111) surface states are formed within a single-unit stripe bounded by SF-induced fractional steps.
Despite an imperfect confinement at the boundaries, the surface states can be described by decoupled 1D states rather than by extended 2D Bloch states.
This may be explained by misalignment of the quantized levels in the transverse direction due to spatially inhomogeneous strains in the films.

Let us briefly mention a possible application of these 1D surface states.
They may be used to mediate electronic and magnetic interactions between surface adsorbates via spatial charge and spin modulations (Friedel/ RKKY oscillations), if the surface states can be tuned to the Fermi level \cite{Silly_2DAtomicSuperlattice,Liu_CoIn4x1,Yin_MagneticCouplingCu(111),Schiffrin_Atom1DBiomolecule}.
This is in principle possible, for example, by substituting the Ag stripes by Au stripes \cite{Puneet_AuAgstripe,Cercellier_AuAgSStuning}.
Since the striped metal surface can be used for self-assembling of organic molecules in the form of a 1D array, they will be utilized for manipulating the spatial and magnetic ordering for organic molecules \cite{Tanaka_CoPcAgStripe}.
Our present study will be the basis for such a forthcoming study.

\begin{acknowledgments}

This work was financially supported by JSPS under KAKENHI Grant No. 21510110. We thank K. Sakamoto for helpful discussions.
\end{acknowledgments}


\begin{references}

\bibitem{Hoegen_AgSi}
M. Horn-von Hoegen, T. Schmidt, G. Meyer, D. Winau, and K. H. Rieder,
	Phys. Rev. B \textbf{52} 10764 (1995).
	
	
\bibitem{Smith_AgGaAs}
A. R. Smith, K.-J. Chao, Q. Niu, and C.-K. Shih,
	Science \textbf{273} 226 (1996).
	
\bibitem{Neuhold_AgSiSS}
G. Neuhold and K. Horn,
	Phys. Rev. Lett. \textbf{78} 1327  (1997).

\bibitem{Gavioli_AgSi}
L. Gavioli, K. R. Kimberlin, M. C. Tringides, J. F. Wendelken, and Z. Zhang,
	Phys. Rev. Lett. \textbf{82} 129 (1999).

\bibitem{Yeh_PbSi}
V. Yeh, L. Berbil-Bautista, C. Z. Wang, K. M. Ho, and M. C. Tringides,
	Phys. Rev. Lett. \textbf{85} 5158 (2000).

\bibitem{Jiang_AgGaAs}
C.-S. Jiang, H.-B. Yu, X. -D. Wang, C.-K. Shih, and Ph. Ebert,
	 Phys. Rev. B \textbf{64} 235410 (2001).

\bibitem{Matsuda_AgSiQWS}
I. Matsuda, H. W. Yeom, T. Tanikawa, K. Tono, T. Nagao, S. Hasegawa, and T. Ohta,
	Phys. Rev. B \textbf{63} 125325 (2001).

\bibitem{Su_PbSi} W. B. Su, S. H. Chang, W. B. Jian, C. S. Chang, L. J. Chen, and T. T. Tsong,
	 Phys. Rev. Lett. \textbf{86} 5116 (2001).

\bibitem{Jiang_AgSi}
C.-S. Jiang, H. Yu, C.-K. Shih, and Ph. Ebert,
	Surf. Sci. \textbf{518} 63 (2002).


\bibitem{Guo_PbQW}
Y. Guo,Y.-F. Zhang, X.-Y. Bao, T.-Z. Han, Z. Tang, L.-X. Zhang, W.-G. Zhu, E. G. Wang, Q. Niu, Z. Q. Qiu, J.-F. Jia, Z.-X. Zhao, Q.-K. Xue,
	Science \textbf{306} 1915 (2004).


\bibitem{Ricci_PbSi} D. A. Ricci, T. Miller, and T.-C. Chiang,
	 Phys. Rev. Lett. \textbf{95} 266101 (2005).

\bibitem{Tang_AgGe}
S.-J. Tang, Y.-R. Lee, S.-L. Chang, T. Miller, and T.-C. Chiang,
	Phys. Rev. Lett. \textbf{96} 216803 (2006). 


\bibitem{Uchihashi_AgStripe}
T. Uchihashi, C. Ohbuchi, S. Tsukamoto, and T. Nakayama,
	Phys. Rev. Lett. \textbf{96} 136104 (2006).


\bibitem{Uchihashi_AgSi(111)In}
T. Uchihashi, T. Nakayama, and M. Aono,
	Jpn. J. Appl. Phys. \textbf{46} 5975 (2007).


\bibitem{Nagamura_AgStripe}
N. Nagamura, I. Matsuda, N. Miyata, T. Hirahara, S. Hasegawa, and T. Uchihashi,
	Phys. Rev. Lett. \textbf{96} 256801 (2006).


\bibitem{Okuda_AgStripes}
T. Okuda, Y. Takeichi, K. He, A. Harasawa, A. Kakizaki, and I. Matsuda,
Phys. Rev. B \textbf{80} 113409 (2009).


\bibitem{Sawa_AgSSDislocation}
K. Sawa, Y. Aoki, and H. Hirayama,
	Phys. Rev. Lett. \textbf{104} 016806 (2010).


\bibitem{Uchihashi_AgSFReflection}
T. Uchihashi, K. Kobayashi, and T. Nakayama,
	 Phys. Rev. B \textbf{82} 113413 (2010).


\bibitem{Abukawa_In4x1}
T. Abukawa, M. Sasaki, F. Hisamatsu, T. Goto, T. Kinoshita, A. Kakizaki, and S. Kono,
Surf. Sci. \textbf{325} 33 (1995).


\bibitem{Kraft_InSurfaces}
J. Kraft, M. G. Ramsey, and F. P. Netzer,
	Phys. Rev. B \textbf{55} (1997) 5384.

\bibitem{Bunk_In4x1}
O. Bunk, G. Falkenberg, J. H. Zeysing, L. Lottermoser, and R. L. Johnson,
	Phys. Rev. B \textbf{59} 12228 (1999).

\bibitem{Yeom_In4x1}
H. W. Yeom, S. Takeda, E. Rotenberg, I. Matsuda, K. Horikoshi, J. Schaefer, C. M. Lee, S. D. Kevan, T. Ohta, T. Nagao, and S. Hasegawa,
	Phys. Rev. Lett. \textbf{82} 4898 (1999).

\bibitem{Nakamura_In4x1}
J. Nakamura, S. Watanabe, and M. Aono,
	Phys. Rev. B \textbf{63} 193307 (2001).

\bibitem{Miwa_In4x1}
R. H. Miwa and G. P. Srivastava,
	Surf. Sci. \textbf{473} 123 (2001).

\bibitem{Kobayashi_SFBulk}
K. Kobayashi and T. Uchihashi,
	Phys. Rev. B \textbf{81} 155418 (2010).

\bibitem{Wisendanger_STMTextbook}
R. Wisendanger, \textit{Scanning Probe Microscopy and Spectroscopy} (Cambridge Univsersity Press, New York, 1994), Sect. 1.13. 


\bibitem{Crommie_QuantumCorral}
M. F. Crommie, C. P. Lutz, and D. M. Eigler,
	Science	\textbf{262} 218 (1993).

\bibitem{Hasegawa_AuSW}
Y. Hasegawa and Ph. Avouris,
	Phys. Rev. Lett. \textbf{71} 1071 (1993).

\bibitem{Avouris_AuQSE}
Ph. Avouris and I.-W. Lyo,
	Science	\textbf{264} (1994) 942.

\bibitem{Burgi_AgResonator}
L. B{\"u}rgi, O. Jeandupeux, A. Hirstein, H. Brune, and K. Kern,
	Phys. Rev. Lett. \textbf{81} 5370 (1998).


\bibitem{Li_AgSW}
	J. Li, W. D. Schneider, and R. Berndt,
	Phys. Rev. B \textbf{56}, 7656 (1997).


\bibitem{Li_AgIsland}	
J. Li, W. D. Schneider, R. Berndt, and S. Crampin,
	Phys. Rev. Lett. \textbf{80} 3332 (1998).

\bibitem{Pivetta_RipplesAgSS}
M. Pivetta, F. Silly, F. Patthey, J. P. Pelz, and W.-D. Schneider,
	Phys. Rev. B \textbf{67}, 193402 (2003).



\bibitem{Pennec_SupramolecularGrating}
Y. Pennec, W. Auw{\"a}rter, A. Schiffrin, A. Weber-Bargioni, A. Riemann, and J. V. Barth,
Nature Nanotech. \textbf{2} 99 (2007). 


\bibitem{Yokoyama_Si(001)SS}
T. Yokoyama, M. Okamoto, and K. Takayanagi,
	Phys. Rev. Lett. \textbf{81} 3423 (1998).

\bibitem{Sagisaka_Si(001)}
K. Sagisaka and D. Fujita,
	 J. Phys.: Conf. Ser. \textbf{100} 052002  (2008).


\bibitem{Morgenstern_AgSSdep}
K. Morgenstern, K.-F. Braun, and K.-H. Rieder,
	Phys. Rev. Lett. \textbf{89} 226801 (2002).

\bibitem{Mugarza_AuResonator}
A. Mugarza, A. Mascaraque, V. P{\'e}rez-Dieste, V. Repain, S. Rousset, F. J. Garc{\'i}a de Abajo, and J. E. Ortega,
	Phys. Rev. Lett. \textbf{87} 107601 (2001).


\bibitem{Ortega_1Dvs2DAgSS}
J. E. Ortega, A. Mugarza, V. Repain, S. Rousset, V. P{\'e}rez-Dieste, and A. Mascaraque,
	Phys. Rev. B \textbf{65} 165413 (2002).

\bibitem{Mugarza_VicinalSurfaces}
A. Mugarza and J. E. Ortega,
	J. Phys.: Condens. Matter \textbf{15} S3281 (2003).


\bibitem{Pendry_1DScatterer}
G. H{\"o}rmandinger and J. B. Pendry,
	Phys. Rev. B \textbf{50}, 18607 (1994).


\bibitem{Ziman_Textbook}
J. M. Ziman, \textit{Principles of the Theory of Solid}  (Cambridge Univsersity Press, New York, 1972) 2nd ed., Sect. 3.4. 


\bibitem{Silly_2DAtomicSuperlattice}
F. Silly, M. Pivetta, M. Ternes, F. Patthey, J. P. Pelz, and W.-D. Schneider,
	Phys. Rev. Lett. \textbf{92} 016101 (2004).

\bibitem{Schiffrin_Atom1DBiomolecule}
A. Schiffrin, J. Reichert, W. Auw\"arter, G. Jahnz, Y. Pennec, A. Weber-Bargioni, V. S. Stepanyuk, L. Niebergall, P. Bruno, and J. V. Barth,
	Phys. Rev. B \textbf{78} 035424 (2008).

\bibitem{Liu_CoIn4x1}
C. Liu, T. Uchihashi, and T. Nakayama,
	Phys. Rev. Lett. \textbf{101} 146104 (2008).


\bibitem{Yin_MagneticCouplingCu(111)}
L. Yin, D. Xiao, Z. Gai, T. Z. Ward, N. Widjaja, G. M. Stocks, Z. Cheng, E. W. Plummer, Z. Zhang, and J. Shen,
	Phys. Rev. Lett. \textbf{104} 167202 (2010).


\bibitem{Cercellier_AuAgSStuning}
H. Cercellier, Y. Fagot-Revurat, B. Kierren, F. Reinert, D. Popovia, and D. Malterre,
Phys. Rev. B \textbf{70} 193412 (2004).


\bibitem{Puneet_AuAgstripe}
P. Mishra, T. Uchihashi, and T. Nakayama,
	Phys. Rev. B \textbf{81} 115430 (2010).

\bibitem{Tanaka_CoPcAgStripe}
Y. Tanaka, P. Mishra, R. Tateishi, H. Orita, M. Otani, T. Uchihashi, and K. Sakamoto 
(unpublished).









\end{references}
\end{document}